# Temporal dynamics of subjective sleepiness: A convergence analysis of two scales

Valeriia Demareva[a]*, Valeriia Viakhireva[a], Irina Zayceva[a], Inna Isakova[a], Yana Okhrimchuk[a], Karina Zueva[a], Andrey Demarev[a], Nikolay Nazarov[a], and Julia Edeleva[a]

[a]*Faculty of Social Sciences, Lobachevsky State University, Nizhny Novgorod, Russia*

Correspondence: valeriia.demareva@fsn.unn.ru

# Temporal dynamics of subjective sleepiness: A convergence analysis of two scales

While sleepiness assessment metrics were initially developed in medical research to study the effects of drugs on sleep, subjective sleepiness assessment is now widely used in both fundamental and applied studies. The Stanford Sleepiness Scale (SSS) and the Karolinska Sleepiness Scale (KSS) are often considered the gold standard in sleepiness research. However, only a few studies have applied both scales, and their convergence and specific features have not been sufficiently investigated. The present study aims to analyse the dynamics and convergence of subjective sleepiness as measured by the KSS and SSS in a population of adults. To achieve this, we present the Subjective Sleepiness Dynamics Dataset (SSDD), which collects evening and morning data on situational subjective sleepiness. A total of 208 adults participated in the experiment. Our findings suggest that sleepiness generally increased from the evening till night and was highest early in the morning. The SSS score appeared to be more sensitive to certain factors, such as the presence of a sleep disorder. The SSS and KSS scores strongly correlated with each other and converged on sleepiness assessment. However, the KSS showed a more even distribution of scores than the SSS. Currently, we are continuously expanding the SSDD.

Keywords: sleepiness; temporal dynamics; convergence; dataset; SSS; KSS

## 1. Introduction

Sleepiness is a natural biological function characterized by a high likelihood of further falling asleep [5]. It is also the tendency to doze off or fall asleep when one intends to stay awake [63] and is often accompanied by a decrease in performance [42]. Sleepiness assessment has been a significant topic in medical research, with early studies focusing on the effects of drugs on sleep [4; 24].

A wide range of methods was employed to study sleep and sleepiness, including physiological and behavioural measures and self-reports. While physiological signals, such as those registered by electroencephalography (EEG) or electrocardiography

(ECG), may appear more robust for detecting sleepiness, understanding sleep as a behaviour remains useful [43]. For example, a subject's performance on behavioural tasks during the process of falling asleep, although highly asynchronized with physiological data, provides a broader understanding of the phase of falling asleep [44]. Models have been proposed that treat falling asleep as a multi-layered and dynamic phenomenon by integrating both physiological measurements and behavioural data to determine sleep coefficient.

Two questionnaires are commonly used to study sleep and sleepiness in relation to sex, age, body weight, and sleep latency [29], as well as the relationship between subjective and objective sleepiness [49]. The Stanford Sleepiness Scale (SSS), released in 1973, is used to quantify and assess progressive stages of situational sleepiness [22]. The SSS was complemented by the Karolinska Sleepiness Scale (KSS) released in 1990 [2]. Another scale, the Epworth Sleepiness Scale (ESS), is a simple, self-administered questionnaire that measures the subject's general (not situational) level of daytime sleepiness [27]. Since it is not the primary focus of the current study, it will not be discussed further. The SSS and the KSS are used in studies of various aspects of sleep duration [21], as well as the influence of both actual [60] and perceived duration [53] on cognitive abilities, sleepiness, and work task performance in people of different ages [7]. Various external factors, such as the effect of light on alertness and sleep, and the sleepiness of drivers on shift schedules, have also been studied [37; 61; 32]. The level of subjective sleepiness on the KSS and the SSS is used to evaluate how cognitive abilities (decision-making, sustained attention, and memory) are affected by factors such as sleep deprivation, restorative sleep, and subjective sleep duration in the general population and in people with epilepsy [45; 1; 38; 46; 56]. A separate group of studies examined age-related changes in various aspects of sleep and sleepiness, such as subjective sleepiness [30], increased age-related tendency to daytime sleep, and decreased slow-wave sleep stage [12], the effect of sleep loss on driving efficiency in people of different ages [7], vision indicators [59], and age-related features of sleepiness response to various substances [51]. The SSS and the KSS were used to assess the effect of caffeine [17; 11], medical drugs potentially causing drowsiness [51], or affecting driving performance [52]. The KSS and the SSS were used in a comprehensive study of apnoea, including the risks of sleep apnoea in patients with an acute stroke [31], the association of apnoea with disease [28], and the effect on alertness while driving [55].

The KSS and the SSS appear to be indispensable in the most voluminous group of studies related to driving characteristics, behaviour, and conditions. Traditionally, these studies focus on the influence of sleepiness and fatigue on driving performance [7; 57; 50; 9; 62], the development of drowsiness detection systems [59; 15; 18; 34; 25] and systems to prevent loss of control due to sleepiness or fatigue [54; 10], the influence of various substances [11; 52], health characteristics [55], or other factors [6; 32] on sleepiness during driving.

Moreover, much attention has been paid to the psychometric properties of these scales [35], such as their sensitivity to short-term partial sleep deprivation and oversleeping [20], as well as cultural differences [26; 36]. The development of new scales has been special focus of research [3]. There is evidence for the dynamics of subjective sleepiness as assessed by the KSS from 19:55 to 23:00 and from 07:05 to 08:10. The sleepiness graph shows an increase in the KSS score from evening to bedtime. At the time of awakening, the KSS score is approximately equal to the one at 19:55 [23]. Additionally, KSS scores for 24-hour monitoring are presented in [40].

In the 1990s, some studies on sleep during driving combined EEG data,

breathing and oxyhemoglobin levels, body temperature, and SSS scores [41]. Another study evaluated the effect of caffeine, short-term sleep, cold air blowing into the face, and listening to music as sleep disturbances, in addition to EEG screening and administration of KSS and Epworth Sleepiness Scale (ESS) [39]. One study found correlations between eye movement metrics, such as saccadic velocity, initial pupil diameter, latency, and amplitude of pupil constriction, and SSS scores [47].

Therefore, the selection of suitable self-report methods for complex studies, which are closely related to the subjective dimension of behavioural methods, remains a problem.

As the KSS and the SSS are often used as the gold standard in articles on sleepiness studies, with very few articles applying both scales [e.g., 14], and practically none studying their congruence and specific features, the present article focuses on the dynamics and correlation of subjective sleepiness as measured by the KSS and the SSS in a population of adults. The objective of this paper is to present a dataset with the evening and morning dynamics of situational subjective sleepiness measured by the KSS and the SSS, as well as sociodemographic characteristics of respondents, including the presence of chronic conditions, caffeine consumption, the presence of sleep disorders, driving experience, sleep and wakefulness patterns, subjective sleep characteristics, and ESS scores. Although heart rate data were recorded (i.e., pulse, RR-intervals, electrocardiogram - ECG), this paper only provides self-report data.

## 2. Materials and methods

### 2.1. Participants

A total of 208 subjects took part in the experiment. The respondents were recruited by circulating the information about the study via available university and regional channels. Potential respondents were invited to fill out a Google form where they indicated their interest to participate in the study as well as their name, sex, age, and contact information. They were then invited to the lab to collect a printed copy of the experiment instruction and equipment (Samsung Smartphone and Polar H10 heart rate sensor). The study design and procedures were approved by the Ethics Committee of Lobachevsky State University, and all participants or their legal representatives provided written informed consent in accordance with the Declaration of Helsinki.

## *2.2. Materials and design*

An automated system was developed to collect and process data. The system was designed as a web service (web application) and implemented in the PHP programming environment using the modern microframework 'CodeIgniter' (version 4). MariaDB was used for data storage.

The system provides several key functionalities:

1. The option for subjects to select a convenient time for testing, collecting, and returning equipment. The system records sensor pickup and drop-off data (based on the number of available sensors) and only allows subjects to sign up for dates when sensors are available for pickup.

2. Online data collection at each testing stage. All experiment steps are programmed sequentially to comply with the required procedures for completing a full test cycle. For cyclic tests (such as the SSS and the KSS), the system automatically blocks the possibility of retaking until the specified amount of time has elapsed.

3. The ability for system administrators to export user test results as a CSV file for further analysis and processing. ESS results can be exported separately.

On the day of the experiment, the experimenter with administrator rights created an account for the participant using their email as the login and a system-generated password in the Web System. The credentials and Web System address were automatically delivered to the participant via email.

The experiment started in the evening of the designated day. At 19:40, participants attached the Polar H10 heart rate sensor and connected it to the Polar Sensor Logger App. Subsequently, at 19:50, participants accessed the Web System and completed the questionnaire as outlined in Appendix 1.

To obtain subjective sleepiness metrics, the original SSS [8] and KSS [2] scales were employed. An additional 10th item ('Extremely sleepy, can't keep awake' [48]) was introduced for the KSS. Participants were instructed to complete the ESS, SSS, and KSS at 20:00. The SSS and KSS were filled out in 30-minute intervals. Participants were informed in advance of the next timestamp when they would need to complete the SSS and KSS. The cycle ceased when participants indicated in the Web System that they were going to bed. A specific time was assigned to this event, which was used for further analysis.

At 06:00, participants completed the KSS, SSS, as well as Levin's subjective sleep characteristics questionnaire [33]. Additionally, they were given the option to provide details about their dreams in a commentary field.

## 3. Results

The data from 15 participants was excluded from the analysis because they had not complied the experiment administration instructions. The common problem was a failure of the participants to observe repeated 30-minute cycles for the SSS and the KSS, which resulted in missing data on subjective sleepiness dynamics.

The remaining sample consisted of 58 males aged 30±11 and 153 females aged 32±12 and was used for statistical analysis. As the primary goal of the study was to estimate the degree of convergence between the SSS and KSS, heart rate and ESS data, as well as subjective sleep characteristics and dream descriptions, were not reported.

Data preprocessing and statistical analysis were performed using the Python programming language with the Jupyter Notebook web-based interactive computing platform. The independent t-test was used to assess differences in SSS and KSS scores at different time points, while the Pearson correlation coefficient was used to assess correlations between the SSS and KSS scores.

*3.1. SSDD descriptive analysis*

Table 1 contains the information about age-sex distribution in the SSDD.

**Table 1 near here**

Figure 1 represents the time points when the cyclic tests (the SSS and the KSS) were completed by each of 191 participants.

**Figure 1 near here**

As shown in Figure 1, all participants completed the SSS and KSS at 20:00 and 06:00. The bedtime of male and female participants did not significantly differ. The most frequent bedtime for women was at 23:00, while males tended to go to bed later at 00:30. However, 23:00 was the second most common bedtime for men.

Due to the variation in the participants' bedtimes (see Figure 2), some participants stopped completing the questionnaires as early as 20:30. Table 2 presents the percentage of incomplete questionnaires for each timepoint during the night.

**Figure 2 near here**

**Table 2 near here**

The text explains that the data in Table 2 shows a 22% loss of data at 22:00, indicating that some subjects had already gone to bed by this time. The cyclic test analysis was based only on the data of the 149 subjects who continued filling out the SSS and the KSS up to 22:00.

Furthermore, Figure 3 shows the distribution of timepoints when the subjects generally wake up, allowing for an assessment of any gradual changes in the dynamics of the SSS and the KSS.

**Figure 3 near here**

*3.2. SSS Scores Analysis*

Figure 4 displays the average, upper and lower quartile values, and outliers for the SSS scores grouped by different factors (Sex, Sleep disorder, Sleep mode comfort, average wakeup time during the week) for each time point. Table 3 provides respective t-values and significance estimations for the SSS scores at each time point for each factor (Sex, Sleep disorder, Sleep mode comfort, average wakeup time during the week).

**Figure 4 near here**

**Table 3 near here**

As shown in Figure 4, the SSS values increase from 20:00 to 06:00. The data in Table 3 indicate that sex did not affect the SSS scores, whereas the presence of sleep disorders, sleep mode comfort, and average wake-up time during the week did affect the SSS scores. Participants with sleep disorders had lower SSS scores at 20:00 ($p = 0.008$) and 20:30 ($p = 0.037$). SSS scores were lower at 20:00 ($p = 0.000$), 20:30 ($p = 0.000$), 21:00 ($p = 0.000$), 21:30 ($p = 0.037$), and 22:00 ($p = 0.014$) for those who reported discomfort with their sleep mode. Participants who usually wake up later than 07:00 had lower SSS scores at 21:30 ($p = 0.015$) and 22:00 ($p = 0.004$).

*3.3. KSS Scores Analysis*

Figure 5 illustrates the mean, upper and lower quartile values, and outliers for the KSS scores divided by various factors (Sex, Sleep disorder, Sleep mode comfort, and average wakeup time during the week) for each time point. The corresponding independent t-test values with respective significance values are presented in Table 4.

**Figure 5 near here**

**Table 4 near here**

As depicted in Figure 5, the KSS scores show an increase from 20:00 to 06:00. Table 4 indicates that sex and sleep disorders did not have a significant effect on the KSS scores, whereas the comfort of sleep mode and the average wakeup time during the week did. Individuals who reported discomfort with their sleep mode had lower KSS scores at 20:00 ($p = 0.004$), 20:30 ($p = 0.004$), 21:00 ($p = 0.002$), and 21:30 ($p = 0.011$). Additionally, participants who typically wake up later than 07:00 had lower KSS scores at 20:30 ($p = 0.050$), 21:30 ($p = 0.027$), and 22:00 ($p = 0.002$).

*3.4. SSS and KSS convergence analysis*

As the SSS and KSS have different score ranges (1-7 for the SSS and 1-10 for the KSS), we separately transformed the scores for each time point using z-scaling. The probability distributions of the SSS and KSS z-scores for different time points are presented in Figure 6.

**Figure 6 near here**

The data presented in Figure 6 demonstrate that the overall distribution patterns of both SSS and KSS z-scores were similar. Further differences will be discussed in the Discussion section.

Table 5 provides Pearson correlation coefficients between the SSS and KSS scores for each time point, indicating a high degree of correlation between the two scales.

**Table 5 near here**

**4. Discussion**

The study's results indicated that sleepiness increased from 20:00 to 22:00 and peaked at 06:00, which is in line with previous findings that subjective sleepiness, as measured by the KSS, was higher in the morning than in the evening [38]. Additionally, previous studies reported that sleepiness was maximal during late evening [1; 50] and early morning [1]. However, our data deviated from the finding that KSS sleepiness was lower in the morning (at 07:00 [23]] and at 08:00 [13]) than during late evening. It is possible that the one-hour difference in our morning measurement at 06:00 may have led to high sleepiness ratings in our study. The discussion section will further explore these differences.

The study results indicate that sex did not have an effect on SSS and KSS scores at any timepoint. Participants with a sleep disorder had lower SSS scores at 20:00 and 20:30, although this factor did not affect KSS scores. Individuals who reported discomfort with their sleep mode had lower SSS scores across all evening timepoints, while KSS scores were lower at 20:00, 20:30, 21:00, and 21:30. Participants who typically wake up later than 07:00 had lower SSS scores at 21:30 and 22:00, whereas KSS scores were lower at 20:30, 21:30, and 22:00. Therefore, the SSS score may be more sensitive to certain factors (e.g., the presence of a sleep disorder). Lower SSS and KSS values during the late evening and at night for individuals who usually go to bed later than 07:00 may be attributed to their late bedtime. Thus, we conclude that individual differences should be considered when analyzing sleepiness, which is consistent with previous research [58].

Overall, the analysis results revealed that both scales converged on the investigated parameters. The correlation analysis indicated a close relationship between the KSS and SSS scores, which confirms the validity of both scales.

However, the analysis of the score distributions revealed some differences between the two scales when using their original non-normalized scales. Across all timepoints, extremely small standardized values were found only for the KSS, which could be explained by the fact that the KSS scale has a larger range without z-standardization, leading participants to provide more varied answers. In contrast, extremely large values were observed for both scales, except for 20:00, where the large extremes occur only for the SSS. Therefore, the larger range of the KSS could not influence the fact that respondents, in general, gave more varied responses concerning extreme points on the scale.

There were differences in the distribution of SSS and KSS scores across timepoints. The z-scores for the SSS were more skewed to the left at 20:00, 20:30, 21:00, and 21:30, while those for the KSS were more evenly distributed. At 22:00, the SSS z-score distribution was skewed to the left, while the KSS z-score distribution was skewed to the right. At 06:00, both z-score distributions were strongly skewed to the right, with a high proportion of high scores. The KSS z-score distribution for this timepoint had a low representation of mean scores (z=0) and a relatively higher representation of extreme values.

Thus, the SSS and the KSS assess the same phenomenon, and in general, the scores on the two scales are congruent. However, the SSS appears to be more sensitive to a wider range of factors. On the other hand, the KSS showed a generally more uniform distribution than the SSS.

**5. Study limitations**

As the purpose of this study was to compare the KSS and the SSS, we had to exclude those subjects who, for some reason, could not complete the cyclic tests according to the instructions. Nonetheless, all the participants completed the KSS and the SSS at the beginning of the experiment (at 20:00). Although we had to exclude 15 participants, their data can still be used for other research purposes since they have provided data for other variables, such as sociodemographic characteristics, chronic illnesses, caffeine consumption, sleep disorders, driving experience, sleep and wakefulness habits, subjective sleep characteristics, the ESS results, heart rate data (pulse, RR-intervals, and ECG), and the KSS and SSS scores at 20:00. Hence, all the 208 recordings can be used for further scientific investigations.

New data are continuously being added to the SSDD. The sample is being balanced by sex, and data from underrepresented age groups, such as older people, are being collected.

**6. Conclusions**

1. The SSDD is a novel dataset that includes subjective sleepiness scores and other characteristics of the participants, such as chronic health conditions, caffeine intake, sleep disorders, driving experience, sleep and wake patterns, subjective sleep quality, ESS scores, and heart rate data (pulse, RR-intervals, and ECG). Currently, the dataset contains data from 208 participants.

2. Sleepiness levels generally increase from evening to night, reaching a maximum in the early morning. Both the SSS and KSS scores follow this pattern.

3. The SSS score may be more sensitive to certain factors, such as the presence of a sleep disorder, than the KSS.

4. The SSS and KSS scores exhibit a strong correlation. However, the KSS scores generally have a more even distribution than the SSS scores.

**Acknowledgments**

This work was supported by the Russian Science Foundation under Grant No. 22-28-20509.

The authors would like to thank all the interns of the Cyberpsychology Lab who communicated with the participants and took part in data collection.

**Declaration of interest statement**

The authors declare no conflict of interest.

**References**

1. Abrahamsen, A., Weihe, P., Debes, F., van Leeuwen, W.M. Sleep, sleepiness, and fatigue on board Faroese fishing vessels. *Nature and science of sleep* **2022**, 14, 347. https://dx.doi.org/10.2147/NSS.S342410
2. Åkerstedt, T., Gillberg, M. Subjective and objective sleepiness in the active individual. *Int J Neurosc* **1990**, 52, 29-37. https://dx.doi.org/10.3109/00207459008994241
3. Bailes, S., Libman, E., Baltzan, M., Amsel, R., Schondorf, R., Fichten, C.S. Brief and distinct empirical sleepiness and fatigue scales. *Journal of Psychosomatic Research* **2006**, 60 (6), 605-613. https://dx.doi.org/10.1016/j.jpsychores.2005.08.015
4. Barmack, J.E. Studies on the psychophysiology of boredom: Part I. The effect of 15 mgs. of benzedrine sulfate and 60 mgs. of ephedrine hydrochloride on blood pressure, report of boredom and other factors. *Journal of Experimental Psychology* **1939**, 25 (5), 494-505. https://dx.doi.org/10.1037/h0054402
5. Bendaoud, I., Etindele Sosso, F.A. Socioeconomic Position and Excessive Daytime Sleepiness: A Systematic Review of Social Epidemiological Studies. *Clocks & Sleep* **2022**, 4, 240-259. https://dx.doi.org/10.3390/clockssleep4020022
6. Bener, A., Yildirim, E., Özkan, T., Lajunen, T. Driver sleepiness, fatigue, careless behavior and risk of motor vehicle crash and injury: Population based case and control study. *Journal of Traffic and Transportation engineering (English edition)* **2017**, 4 (5), 496-502. https://dx.doi.org/10.1016/j.jtte.2017.07.005
7. Cai, A.W., Manousakis, J.E., Singh, B., Kuo, J., Jeppe, K.J., Francis-Pester, E., Anderson, C. On-road driving impairment following sleep deprivation differs according to age. *Scientific reports* **2021**, 11 (1), 1-13. https://dx.doi.org/10.1038/s41598-021-99133-y
8. Carskadon, M.A., Dement, W.C. Sleepiness and Sleep State on a 90-Min Schedule. *Psychophysiology* **1977**, 14 (2), 127-133. https://dx.doi.org/10.1111/j.1469-8986.1977.tb03362.x
9. Chai, R., Ling, S.H., San, P.P., Naik, G.R., Nguyen, T.N., Tran, Y., Ashley, C., Hung, T., Nguyen, H.T. Improving EEG-based driver fatigue classification using sparse-deep belief networks. *Frontiers in neuroscience* **2017**, 11, 103. https://dx.doi.org/10.3389/fnins.2017.00103

10. Chang, Y.L., Feng, Y.C., Chen, O.T.C. Real-time physiological and facial monitoring for safe driving. In Proceedings of the 38th Annual International Conference of the IEEE Engineering in Medicine and Biology Society (EMBC). 16-20 August 2016. pp. 4849-4852. https://dx.doi.org/10.1109/EMBC.2016.7591813

11. De Valck, E., De Groot, E., Cluydts, R. Effects of slow-release caffeine and a nap on driving simulator performance after partial sleep deprivation. *Perceptual and motor skills* **2003**, 96 (1), 67-78. https://dx.doi.org/10.2466/pms.2003.96.1.67

12. Dijk, D.J., Groeger, J.A., Stanley, N., Deacon, S. Age-related reduction in daytime sleep propensity and nocturnal slow wave sleep. *Sleep* **2010**, 33 (2), 211-223. https://dx.doi.org/10.1093/sleep/33.2.211

13. Flaa, T.A., Bjorvatn, B., Pallesen, S., Zakariassen, E., Harris, A., Gatterbauer-Trischler P., Waage S. Sleep and sleepiness measured by diaries and actigraphy among Norwegian and Austrian helicopter emergency medical service (HEMS) pilots. *International journal of environmental research and public health* **2022**, 19 (7), 4311. https://dx.doi.org/10.3390/ijerph19074311

14. Ganesan, S., Magee, M., Stone, J.E., Mulhall, M.D., Collins, A., Howard, M.E., Sletten, T.L. The impact of shift work on sleep, alertness and performance in healthcare workers. Scientific reports 2019, 9 (1), 4635. https://dx.doi.org/10.1038/s41598-019-40914- x

15. Gaspa, J.G., Brown, T.L., Schwarz, C.W., Lee, J.D., Kang, J., Higgins, J.S. Evaluating driver drowsiness countermeasures. *Traffic injury prevention* **2017**, 18 (sup1), S58-S63. https://dx.doi.org/10.1080/15389588.2017.1303140

16. Gorgoni, M., Scarpelli, S., Annarumma, L., D'atri A., Alfonsi V., Ferrara M., De Gennaro L. The Regional EEG Pattern of the Sleep Onset Process in Older Adults. *Brain Sciences* **2021**, 11 (10), 1261. https://dx.doi.org/10.3390/brainsci11101261

17. Hansen, D.A., Ramakrishnan, S., Satterfield, B.C., Wesensten, N.J., Layton, M.E., Reifman, J., Van Dongen, H. Randomized, double-blind, placebo-controlled, crossover study of the effects of repeated-dose caffeine on neurobehavioral performance during 48 h of total sleep deprivation.

*Psychopharmacology* **2019**, 236 (4), 1313-1322. https://dx.doi.org/10.1007/s00213-018-5140-0

18. He, J., Choi, W., Yang, Y., Lu, J., Wu, X., Peng, K. Detection of driver drowsiness using wearable devices: A feasibility study of the proximity sensor. *Applied ergonomics* **2017**, 65, 473-480. https://dx.doi.org/10.1016/j.apergo.2017.02.016

19. He, J., Choi, W., Yang, Y., Lu, J., Wu, X., Peng, K. Detection of driver drowsiness using wearable devices: A feasibility study of the proximity sensor. *Applied ergonomics* **2017**, 65, 473-480. https://dx.doi.org/10.1016/j.apergo.2017.02.016

20. Herscovitch, J., Broughton, R. Sensitivity of the Stanford sleepiness scale to the effects of cumulative partial sleep deprivation and recovery oversleeping. *Sleep* **1981**, 4 (1), 83-92. https://dx.doi.org/10.1093/sleep/4.1.83

21. Hilditch, C.J., Centofanti, S.A., Dorrian, J., Banks, S. A 30-minute, but not a 10-minute nighttime nap is associated with sleep inertia. *Sleep* **2016**, 39 (3), 675-685. https://dx.doi.org/10.5665/sleep.5550

22. Hoddes, E., Zarcone, V., Smythe, H., Phillips, R., Dement, W.C. Quantification of Sleepiness: A New Approach. *Psychophysiology* **1973**, 10 (4), pp. 431-436. https://dx.doi.org/10.1111/j.1469-8986.1973.tb00801.x

23. Höhn, C., Schmid, S.R., Plamberger, C.P., Bothe, K., Angerer, M., Gruber, G., Pletzer, B., Hoedlmoser, K. Preliminary Results: The Impact of Smartphone Use and Short-Wavelength Light during the Evening on Circadian Rhythm, Sleep and Alertness. *Clocks&Sleep* **2021**, 3, 66–86. https://dx.doi.org/10.3390/clockssleep3010005

24. Hollister, L.E., Clyde, D.J. Blood levels of pentobarbital sodium, meprobamate, and tybamate in relation to clinical effects. *Clinical Pharmacology & Therapeutics* **1968**, 9 (2), 204-208. https://dx.doi.org/10.1002/cpt196892204

25. Hu, X., Eberhart, R., Foresman, B. Modeling drowsy driving behaviors. In Proceedings of 2010 IEEE International Conference on Vehicular Electronics and Safety. 15-17 July 2010. pp. 13-17. https://dx.doi.org/10.1109/ICVES.2010.5550949

26. Izquierdo-Vicario, Y., Ramos-Platon, M.J., Conesa-Peraleja, D., Lozano-Parra, A.B., Espinar-Sierra, J. Letter to the editor. Epworth Sleepiness Scale in a

Sample of the Spanish Population. *Sleep* **1997**, 20 (8), 676-677. https://dx.doi.org/10.1093/sleep/20.8.676

27. Johns, M.W. A new method for measuring daytime sleepiness: The Epworth sleepiness scale. *Sleep* **1991**, 14 (6), 540-545. https://dx.doi.org/10.1093/sleep/14.6.540

28. Kang, H.H., Lim, C.H., Oh, J.H., Cho, M.J., Lee, S.H. The influence of gastroesophageal reflux disease on daytime sleepiness and depressive symptom in patients with obstructive sleep apnea. *Journal of Neurogastroenterology and Motility* **2021**, 27 (2), 215. https://dx.doi.org/10.5056/jnm20071

29. Kim, H., Young, T. Subjective daytime sleepiness: dimensions and correlates in the general population. *Sleep* **2005**, 28(5), 625-634. https://dx.doi.org/10.1093/SLEEP/28.5.625

30. Kim, H., Young, T. Subjective daytime sleepiness: dimensions and correlates in the general population. *Sleep* **2005**, 28 (5), 625-634. https://dx.doi.org/10.1093/SLEEP/28.5.625

31. Kojic, B., Dostovic, Z., Ibrahimagic, O.C., Smajlovic, D., Hodzic, R., Iljazovic, A., Salihovic, D. Risk Factors in Acute Stroke Patients With and Without Sleep Apnea. *Medical Archives* **2021**, 75 (6), 444. https://dx.doi.org/10.5455/medarh.2021.75.444-450

32. Leger, D., Philip, P., Jarriault, P., Metlaine, A., Choudat, D. Effects of a combination of napping and bright light pulses on shift workers' sleepiness at the wheel: a pilot study. *Journal of sleep research* **2009**, 18 (4), 472-479. https://dx.doi.org/10.1111/j.1365-2869.2008.00676.x

33. Levin, Ia.I., Eligulashvili, T.S., Posokhov. S.I., Kovrov, G.V., Bashmakov, M.Y. Farmakoterapiia insomnii: rol' Imovana. In *Sleep disorders*; Aleksandrovskii Iu.A., Vein A.M. Eds.; Med. inform. Agentstvo: Saint Peersburgh, Russia, 1995, pp. 56–61.

34. Li, R., Chen, Y.V., Zhang, L. A method for fatigue detection based on Driver's steering wheel grip. *International Journal of Industrial Ergonomics* **2021**, 82, 103083. https://dx.doi.org/10.1016/j.ergon.2021.103083

35. Maclean, A.W., Fekken, G.C., Saskin, P., Knowles, J.B. Psychometric evaluation of the Stanford sleepiness scale. *Journal of Sleep Research* **1992**, 1 (1), 35-39. https://dx.doi.org/10.1111/j.1365-2869.1992.tb00006.x

36. Malencia-Flores, M., Rosenthal, L., Castaño, V.A., Campos, R.M., Vergara, P., Resendiz, M., Javier Aguilar, R.A.-R., Bliwise, D.L. A factor replication of the Sleep-Wake Activity Inventory (SWAI) in a Mexican population. *Sleep* **1997**, 20 (2), 111-114. https://dx.doi.org/10.1093/sleep/20.2.111

37. Mason, I.C., Grimaldi, D., Reid, K.J., Warlick, C.D., Malkani, R.G., Abbott, S. M., Zee, P.C. Light exposure during sleep impairs cardiometabolic function. *Proceedings of the National Academy of Sciences* **2022**, 119 (12), e2113290119. https://dx.doi.org/10.1073/pnas.2113290119

38. Maurer, L., Zitting, K.M., Elliott, K., Czeisler, C.A., Ronda, J.M., Duffy, J.F. A new face of sleep: the impact of post-learning sleep on recognition memory for face-name associations. *Neurobiology of learning and memory* **2015**, 126. P. 31-38. https://dx.doi.org/10.1016/j.nlm.2015.10.012

39. Maycock, G. Sleepiness and driving: The experience of UK car drivers. *Journal of Sleep Research* **1996**, 5 (4), 229-231. https://dx.doi.org/10.1111/j.1365-2869.1996.00229.x

40. Miley, A.Å., Kecklund, G., Åkerstedt, T. Comparing two versions of the Karolinska Sleepiness Scale (KSS). *Sleep Biol Rhythms* **2016**, 14 (3), 257-260. https://dx.doi.org/10.1007/s41105-016-0048-8

41. Miller, J.C. Batch processing of 10000 h of truck driver EEG data. *Biological Psychology* **1995**, 40 (1-2), 209-222. https://dx.doi.org/10.1016/0301-0511(95)05114-7

42. Muck, R.A., Hudson, A.N., Honn, K.A., Gaddameedhi, S., Van Dongen, H.P.A. Working around the Clock: Is a Person's Endogenous Circadian Timing for Optimal Neurobehavioral Functioning Inherently Task-Dependent? *Clocks&Sleep* **2022**, 4, 23–36. https://dx.doi.org/10.3390/clockssleep4010005

43. Ogilvie, R.D., Simons, I.A., Kuderian, R.H., MacDonald, T., Rustenburg, J. Behavioral, event-related potential, and EEG/FFT changes at sleep onset. *Psychophysiology* **1991**, 28 (1), 54-64. https://dx.doi.org/10.1111/j.1469-8986.1991.tb03386.x

44. Prerau, M.J., Hartnack, K.E., Obregon-Henao, G., Sampson, A., Merlino, M., Gannon, K., Matt, T.B., Jeffrey, M., Purdon, P.L. Tracking the sleep onset process: an empirical model of behavioral and physiological dynamics. *PLoS computational biology* **2014**, 10 (10), e1003866. https://dx.doi.org/10.1371/journal.pcbi.1003866.

45. Rabat, A., Gomez-Merino, D., Roca-Paixao, L., Bougard, C., Van Beers, P., Dispersyn, G., Chennaoui, M. Differential kinetics in alteration and recovery of cognitive processes from a chronic sleep restriction in young healthy men. *Frontiers in Behavioral Neuroscience* **2016**, 10, 95. https://dx.doi.org/10.3389/fnbeh.2016.00095

46. Rahman, S.A., Rood, D., Trent, N., Solet, J., Langer, E.J., Lockley, S.W. Manipulating sleep duration perception changes cognitive performance–an exploratory analysis. *Journal of psychosomatic research* **2020**, 132, 109992. https://dx.doi.org/10.1016/j.jpsychores.2020.109992

47. Russo, M., Thomas, M., Thorne, D., Sing, H., Redmond, D., Rowland, L., Johnson, D., Hall, S., Krichmar, J., Balkin, T. Oculomotor impairment during chronic partial sleep deprivation. *Clinical Neurophysiology* **2003**, 114 (4), 723-736. https://dx.doi.org/10.1016/s1388-2457(03)00008-7

48. Shahid, A., Wilkinson, K., Marcu, S., Shapiro, C.M. Karolinska Sleepiness Scale (KSS). In: Shahid, A., Wilkinson, K., Marcu, S., Shapiro, C. (eds) *STOP, THAT and One Hundred Other Sleep Scales*. Springer: New York, USA, 2011. https://dx.doi.org/10.1007/978-1-4419-9893-4_47

49. Short, M., Lack, L., Wright, H. Does subjective sleepiness predict objective sleep propensity? *Sleep* **2010**, 33 (1), 123-129. https://dx.doi.org/10.1093/sleep/33.1.123

50. Smith, S.S., Horswill, M.S., Chambers, B., Wetton, M. Hazard perception in novice and experienced drivers: The effects of sleepiness. *Accident Analysis and Prevention* **2009**, 41 (4), 729-733. https://dx.doi.org/10.1016/j.aap.2009.03.016

51. Staskin, D.R., Harnett, M.D. Effect of trospium chloride on somnolence and sleepiness in patients with overactive bladder. *Current Urology Reports* **2004**, 5 (6), 423-426. https://dx.doi.org/10.1007/s11934-004-0064-0

52. Suhner, A., Schlagenhauf, P., Tschopp, A., Hauri-Bionda, R., Friedrich-Koch, A., Steffen, R. Impact of melatonin on driving performance. *Journal of travel medicine* **1998**, 5 (1), 7-13. https://dx.doi.org/10.1111/j.1708-8305.1998.tb00448.x

53. Šušmáková, K. Correlation dimension versus fractal exponent during sleep onset. *Measurement science review* **2006**, 6 (4), 58-62.

54. Takayama, L., Nass, C. Assessing the effectiveness of interactive media in improving drowsy driver safety. *Human factors* **2008**, 50(5), 772-781. https://dx.doi.org/10.1518/001872008X312341
55. Tippin, J., Sparks, J.D., Rizzo, M. Visual vigilance in drivers with obstructive sleep apnea. *Journal of psychosomatic research* **2009**, 67 (2), 143-151. https://dx.doi.org/10.1016/j.jpsychores.2009.03.015
56. Vascouto, H.D., Thais, M.E.R.D.O., Osório, C.M., Ben, J., Claudino, L.S., Hoeller, A.A., Hans, J.M., Wolf, P., Lin, K., Walz, R. Is self-report sleepiness associated with cognitive performance in temporal lobe epilepsy? *Arquivos de Neuro-Psiquiatria* **2018**, 76, 575-581. https://dx.doi.org/10.1590/0004-282X20180089
57. Wang, L., Pei, Y. The impact of continuous driving time and rest time on commercial drivers' driving performance and recovery. *Journal of safety research* **2014**, 50, 11-15. https://dx.doi.org/10.1016/j.jsr.2014.01.003
58. Wang, X., Xu, C. Driver drowsiness detection based on non-intrusive metrics considering individual specifics. *Accident Analysis and Prevention* **2016**, 95, 350-357. https://dx.doi.org/10.1016/j.aap.2015.09.002
59. Wang, Y., Xin, M., Bai, H., Zhao, Y. Can variations in visual behavior measures be good predictors of driver sleepiness? A real driving test study. *Traffic injury prevention* **2017**, 18 (2), 132-138. https://dx.doi.org/10.1080/15389588.2016.1203425
60. Yamazaki, E.M., Antler, C.A., Lasek, C.R., Goel, N. Residual, differential neurobehavioral deficits linger after multiple recovery nights following chronic sleep restriction or acute total sleep deprivation. *Sleep* **2021**, 44 (4), zsaa224. https://dx.doi.org/10.1093/sleep/zsaa224
61. Yang, M., Ma, N., Zhu, Y., Su, Y.C., Chen, Q., Hsiao, F.C., Zhou, G. The acute effects of intermittent light exposure in the evening on alertness and subsequent sleep architecture. *International journal of environmental research and public health* **2018**, 15 (3), 524. https://dx.doi.org/10.3390/ijerph15030524
62. Yao, Y., Zhao, X., Du, H., Zhang, Y., Zhang, G., Rong, J. Classification of fatigued and drunk driving based on decision tree methods: a simulator study. *International journal of environmental research and public health* **2019**, 16 (11), 1935. https://dx.doi.org/10.3390/ijerph16111935

63. Yu, Y.-K., Yao, Z.-Y., Wei, Y.-X., Kou, C.-G., Yao, B., Sun, W.-J., Li, S.-Y., Fung, K., Jia, C.-X. Depressive Symptoms as a Mediator between Excessive Daytime Sleepiness and Suicidal Ideation among Chinese College Students. *Int. J. Environ. Res. Public Health* **2022**, 19, 16334. https://dx.doi.org/10.3390/ijerph192316334

**Appendices**

Appendix 1. Questionnaire in the Web System

- Identify your sex.
- Identify your age.
- Identify your height.
- Identify your weight.
- Do you have any chronic diseases. If yes, which ones?
- Do you drive a car regularly?
- If yes, what is your driving record?
- How many cups of coffee do you usually drink per day?
- How many cups of coffee have you had today?
- Do you smoke?
- How long have you been smoking?
- Do you classify yourself as:
    - A night owl
    - An early bird
    - Cannot decide
- What time do you usually get up on weekdays?
- What time do you usually get up on weekends?
- What time do you usually go to bed on weekdays?
- What time do you usually go to bed on weekends?
- Does your sleep pattern meet your current resting needs?
- Do you have any sleep disorders? If so, describe.

**Tables**

Table 1. The number of males and women per age range.

| Age range | Males | Females |
|:---:|:---:|:---:|
| 25-35 | 18 | 34 |
| < 25 | 21 | 45 |
| > 35 | 19 | 54 |
| Total | 58 | 133 |

Table 2. Missing data ratio for each time point.

| № | Timepoint | Missing data ratio | № | Timepoint | Missing data ratio |
|---|-----------|--------------------|---|-----------|--------------------|
| 1 | 20:00 | 0% | 9 | 0:00 | 80% |
| 2 | 20:30 | 4% | 10 | 0:30 | 87% |
| 3 | 21:00 | 6% | 11 | 1:00 | 92% |
| 4 | 21:30 | 12% | 12 | 1:30 | 96% |
| 5 | 22:00 | 22% | 13 | 2:00 | 98% |
| 6 | 22:30 | 32% | 14 | 2:30 | 99% |
| 7 | 23:00 | 52% | 15 | 3:00 | 99% |
| 8 | 23:30 | 65% | 16 | 6:00 | 0% |

Table 3. Independent t-test and *p*-values comparing the SSS scores according to different factors (Sex, Sleep disorder, Sleep mode comfort, general wakeup time during the week) for each time point. Significant differences are shaded in grey ($p < 0.05$).

| Timepoint | 20:00 | 20:30 | 21:00 | 21:30 | 22:00 | 06:00 |
|---|---|---|---|---|---|---|
| **Sex (males vs females)** | | | | | | |
| t | -1.18 | -1.14 | -0.49 | -0.27 | -1.17 | -1.55 |
| p | 0.241 | 0.257 | 0.625 | 0.79 | 0.245 | 0.124 |
| **Sleep disorder (yes vs no)** | | | | | | |
| t | 2.7 | 2.1 | 1.37 | 1.11 | 0.67 | -1.79 |
| p | 0.008 | 0.037 | 0.172 | 0.268 | 0.504 | 0.076 |
| **Sleep mode comfort (yes vs no)** | | | | | | |
| t | -3.93 | -3.87 | -3.63 | -2.1 | -2.5 | 0.14 |
| p | 0.000 | 0.000 | 0.000 | 0.037 | 0.014 | 0.888 |
| **Weekly general wakeup (later than 7:00 vs earlier than 7:00)** | | | | | | |
| t | 1.16 | -1.89 | -1.64 | -2.46 | -2.92 | 1.77 |
| p | 0.25 | 0.061 | 0.103 | 0.015 | 0.004 | 0.079 |

Table 4. Independent t-test and *p*-values comparing the KSS scores according to different factors (Sex, Sleep disorder, Sleep mode comfort, average wakeup time during the week) for each time point. Significant differences are shaded in grey (p < 0.05).

| Time point | 20:00 | 20:30 | 21:00 | 21:30 | 22:00 | 06:00 |
|---|---|---|---|---|---|---|
| **Sex (males vs females)** | | | | | | |
| t | -1.49 | -0.45 | -0.59 | 0.19 | -1.12 | -1.26 |
| p | 0.139 | 0.652 | 0.558 | 0.847 | 0.265 | 0.208 |
| **Sleep disorder (yes vs no)** | | | | | | |
| t | 1.22 | 1.31 | 1.36 | 0.63 | 0.64 | -1.24 |
| p | 0.226 | 0.193 | 0.177 | 0.527 | 0.526 | 0.218 |
| **Sleep mode comfort (yes vs no)** | | | | | | |
| t | -2.94 | -2.93 | -3.21 | -2.59 | -1.72 | -0.99 |
| p | 0.004 | 0.004 | 0.002 | 0.011 | 0.088 | 0.325 |
| **Weekly general wakeup (later than 7:00 vs earlier than 7:00)** | | | | | | |
| t | 0.02 | -1.98 | -1.5 | -2.23 | -3.12 | 1.41 |
| p | 0.982 | 0.050 | 0.136 | 0.027 | 0.002 | 0.162 |

Table 5. Pearson correlation coefficient values (R) between the SSS and the KSS scores, lower and higher confidence intervals (CI), and *p*-values for each timepoint.

| Timepoint | R | Lower CI (p = 0.05) | Upper CI (p = 0.05) | p |
|---|---|---|---|---|
| 20:00 | 0.75 | 0.67 | 0.81 | 0.00000 |
| 20:30 | 0.82 | 0.76 | 0.87 | 0.00000 |
| 21:00 | 0.82 | 0.76 | 0.86 | 0.00000 |
| 21:30 | 0.81 | 0.74 | 0.86 | 0.00000 |
| 22:00 | 1.00 | 0.79 | 0.88 | 0.00000 |
| 06:00 | 0.89 | 0.85 | 0.92 | 0.00000 |

# Figures

Figure 1

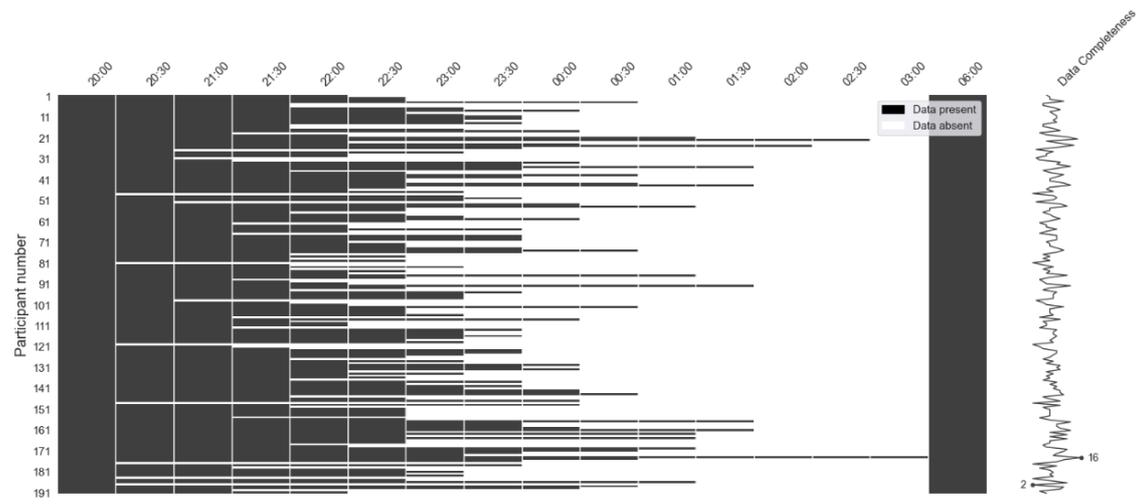

Figure 2

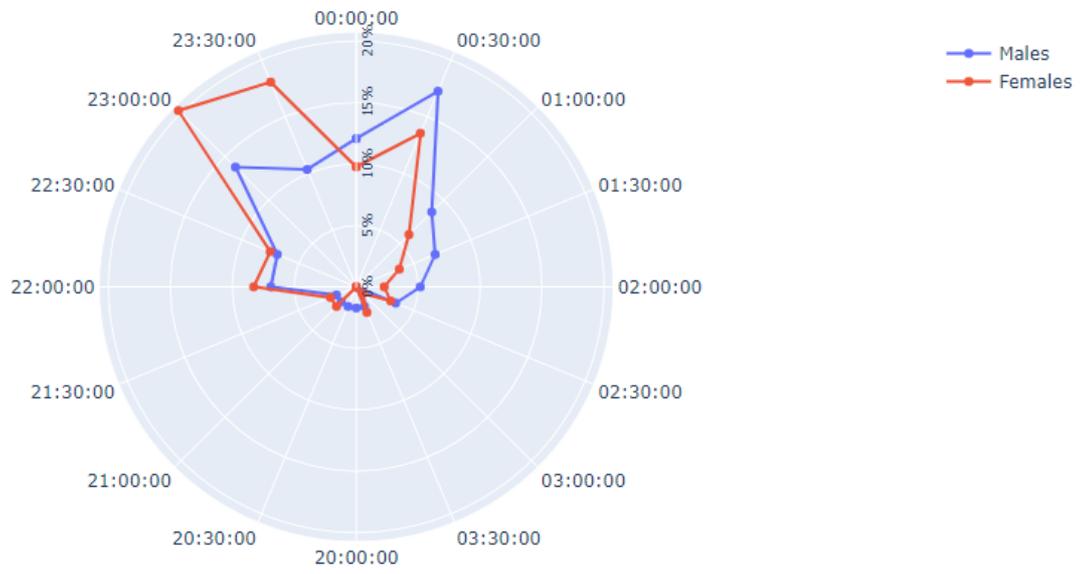

Figure 3

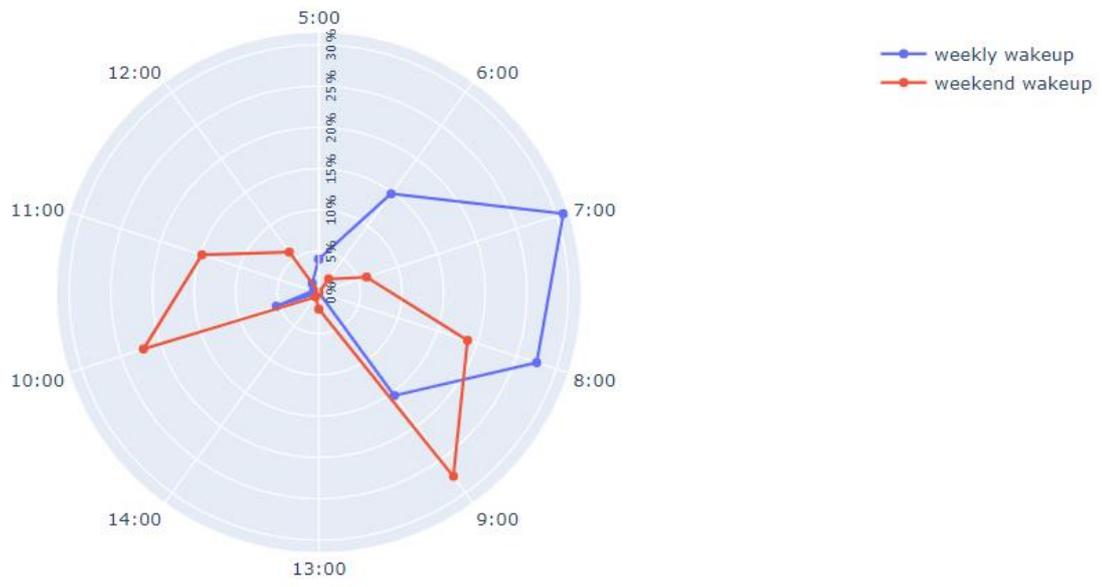

Figure 4

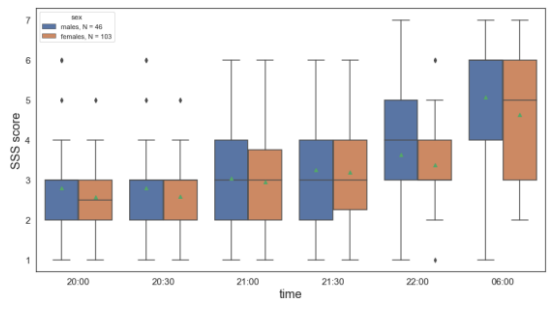

(a)

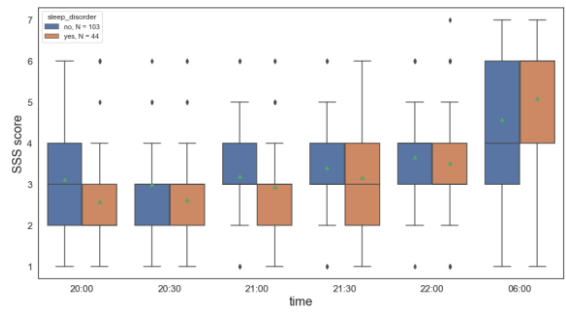

(b)

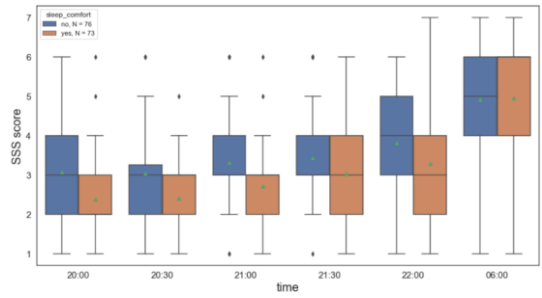

(c)

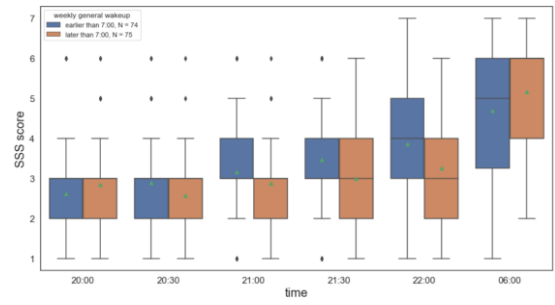

(d)

Figure 5

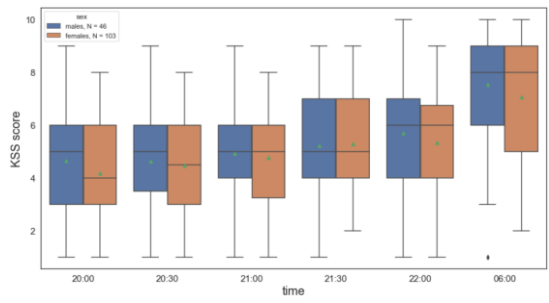

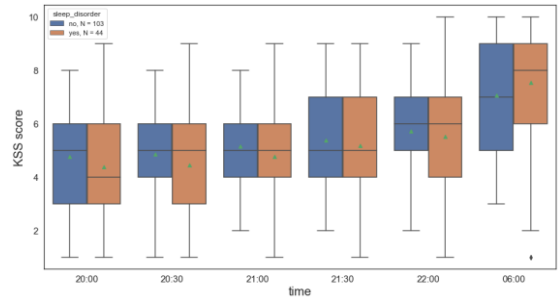

(a)

(b)

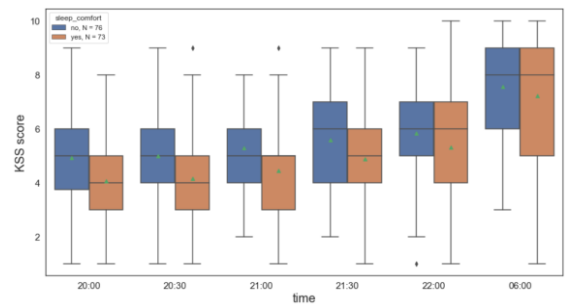

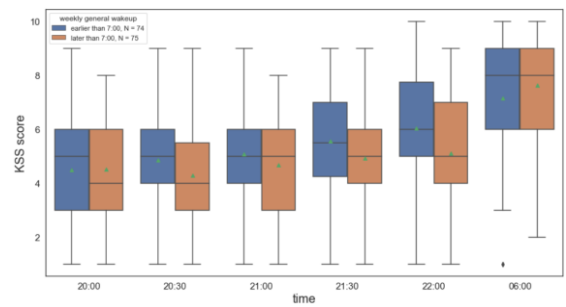

(c)

(d)

Figure 6

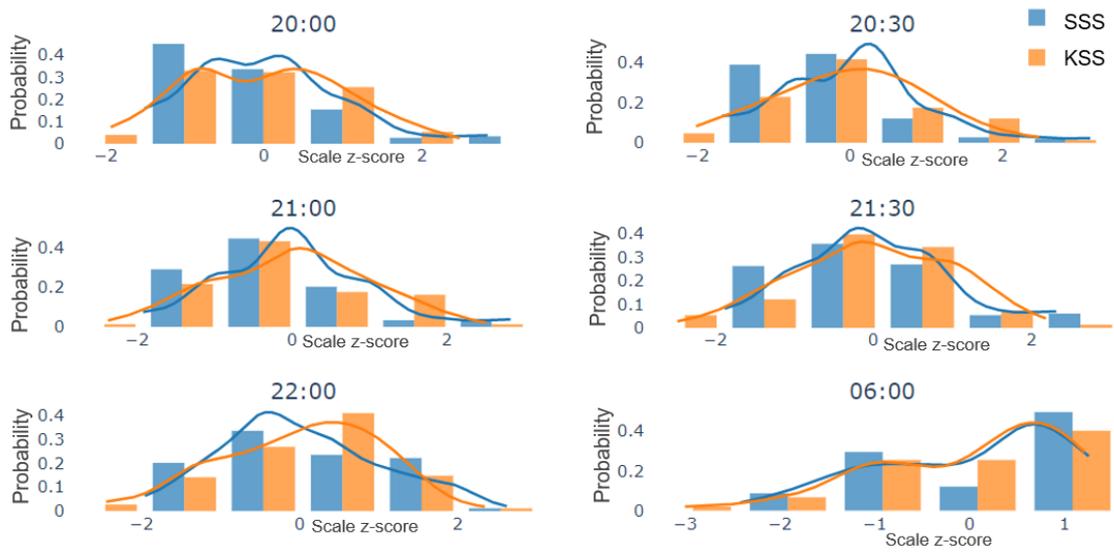

**Figure captions**

Figure 1. The incidence of missing values and data completeness.

Figure 2. The distribution of the times the subjects (males and females) went to bed at the day of experiment.

Figure 3. The distribution of average wake up time during the week and at weekends.

Figure 4. The distribution of the SSS values grouped by different factors ((a) Sex, (b) Sleep disorder, (c) Sleep mode comfort, (d) Average wakeup time during the week) for each time point. Green triangles represent mean values.

Figure 5. The distribution of the KSS values grouped by different factors ((a) Sex, (b) Sleep disorder, (c) Sleep mode comfort, (d) average wakeup time during the week) for each time point. Green triangles represent mean values.

Figure 6. Probability distributions for the SSS and KSS z-scores for each timepoint. Lines indicate kernel density estimates.